\documentclass[manuscript]{acmart}

\AtBeginDocument{%
  \providecommand\BibTeX{{%
    \normalfont B\kern-0.5em{\scshape i\kern-0.25em b}\kern-0.8em\TeX}}}

\setcopyright{acmlicensed}
\copyrightyear{2025}
\acmYear{2025}
\acmDOI{}

\newboolean{showcomments}
\setboolean{showcomments}{true} % toggle to show or hide comments
\ifthenelse{\boolean{showcomments}}
{\newcommand{\nb}[2]{
		\fcolorbox{black}{yellow}{\bfseries\sffamily\scriptsize#1}
		{\sf\small$\blacktriangleright$\textit{#2}$\blacktriangleleft$}
	}
	
}
{\newcommand{\nb}[2]{}
	
}

\usepackage{balance}
\usepackage{graphicx,subcaption,lipsum}
\usepackage{array}
\setcopyright{none}

\begin{document}

\title{See What I See: An Attention-Guiding eHMI Approach for Autonomous Vehicles}

\author{Jialong Li}
\affiliation{%
  \institution{Waseda University}
  \city{Tokyo}
  \country{Japan}}
\email{lijialong@fuji.waseda.jp}

\author{Zhiyao Wang}
\affiliation{%
  \institution{The University of Osaka}
  \city{Osaka}
  \country{Japan}}
\email{wangzhiyao@ist.osaka-u.ac.jp}

\author{Zhenyu Mao}
\affiliation{%
  \institution{City University of Hong Kong}
  \city{Hong Kong}
  \country{China}}
\email{zhenyumao2-c@my.cityu.edu.hk}

\author{Yijun Lu}
\affiliation{%
  \institution{Waseda University}
  \city{Tokyo}
  \country{Japan}}
\email{yijun@ruri.waseda.jp}

\author{Shogo Morita}
\affiliation{%
  \institution{Institute of Science Tokyo}
  \city{Tokyo}
  \country{Japan}}
\email{morita.s.8286@m.isct.ac.jp}

\author{Nianyu Li}
% \authornote{Corresponding Author: }
\affiliation{%
  \institution{ZGC Laboratory}
  \city{Beijing}
  \country{China}}
\email{li\_nianyu@pku.edu.cn}

\author{Kenji Tei}
\affiliation{%
  \institution{Institute of Science Tokyo}
  \city{Tokyo}
  \country{Japan}}
\email{tei@comp.isct.ac.jp}

\renewcommand{\shortauthors}{J Li. et al.}

\begin{abstract}
As autonomous vehicles are gradually being deployed in the real world, external Human--Machine Interfaces (eHMIs) are expected to serve as a critical solution for enhancing vehicle--pedestrian communication. However, existing eHMI designs typically focus solely on the ego vehicle’s status, which can inadvertently capture pedestrians’ attention or encourage misguided reliance on the AV’s signals, leading them to neglect scanning for other surrounding hazards.
To address this, we propose the Attention-Guiding eHMI (AGeHMI), a projection-based visualization that employs directional cues and risk-based color coding to actively guide pedestrians' attention toward potential environmental dangers. Evaluation through a virtual reality user study ($N = 20$) suggests that AGeHMI effectively influences participants' visual attention distribution and significantly reduces potential collision risks with surrounding vehicles, while simultaneously improving subjective confidence and reducing cognitive workload.
\end{abstract}




\maketitle
% \vspace{-0.5cm}

\section{Introduction}
\vspace{-2mm}
\begin{figure*}[h!]
    \centering
    % \vspace{-1.5mm}
    \begin{subfigure}[t]{0.28\linewidth}
        \centering
        \includegraphics[width=\linewidth]{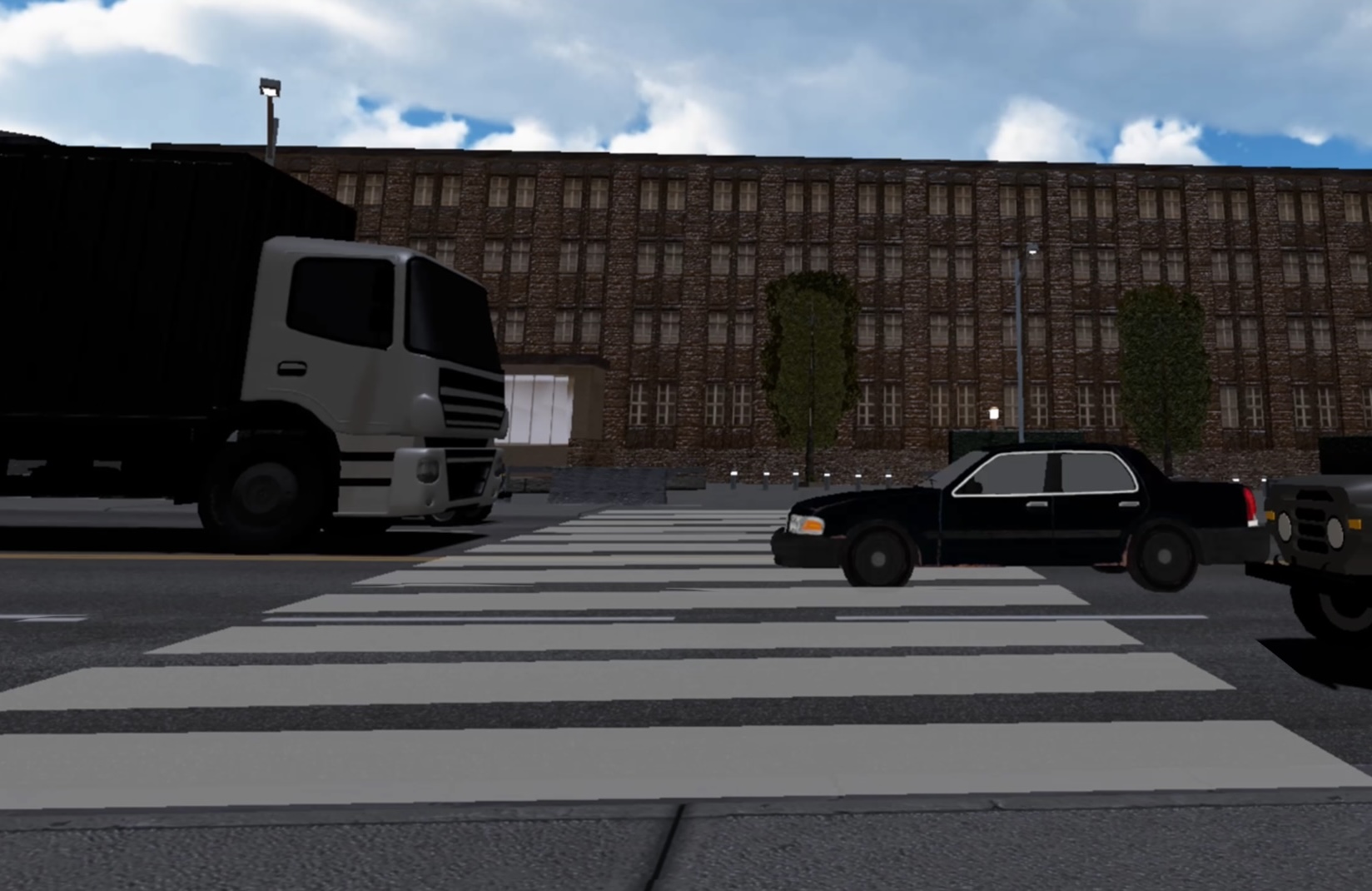}
        \caption{\centering Baseline without eHMI}
        \label{fig:no_ehmi}
    \end{subfigure}
    \hfill
    \begin{subfigure}[t]{0.34\linewidth}
        \centering
        \includegraphics[width=\linewidth]{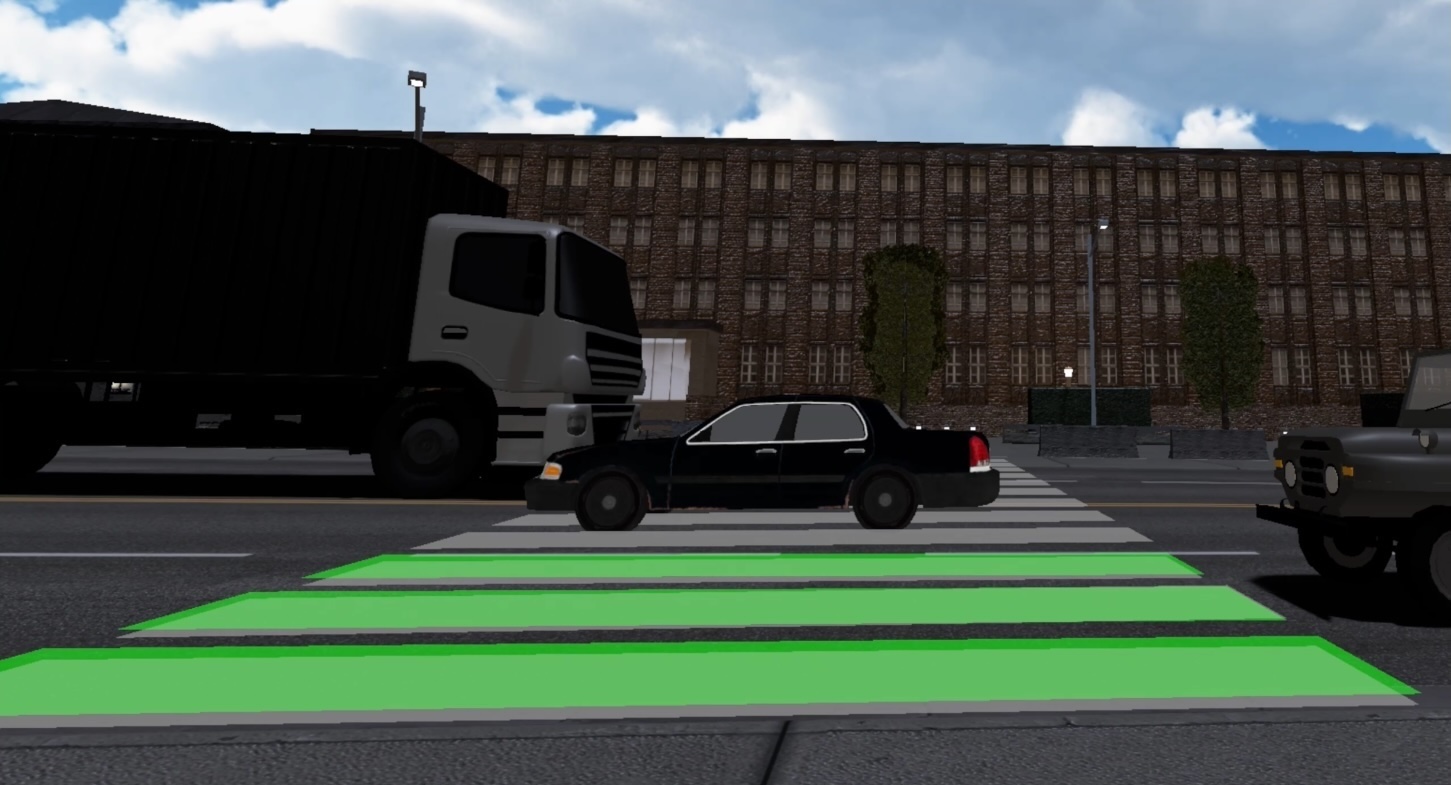}
        % \caption{\centering eHMI displaying ego-vehicle intent only}
        \caption{\centering eHMI in related studies\\ convey ego vehicle's intent only}
        \label{fig:related_ehmi}
    \end{subfigure}
    \hfill
    \begin{subfigure}[t]{0.34\linewidth}
        \centering
        \includegraphics[width=\linewidth]{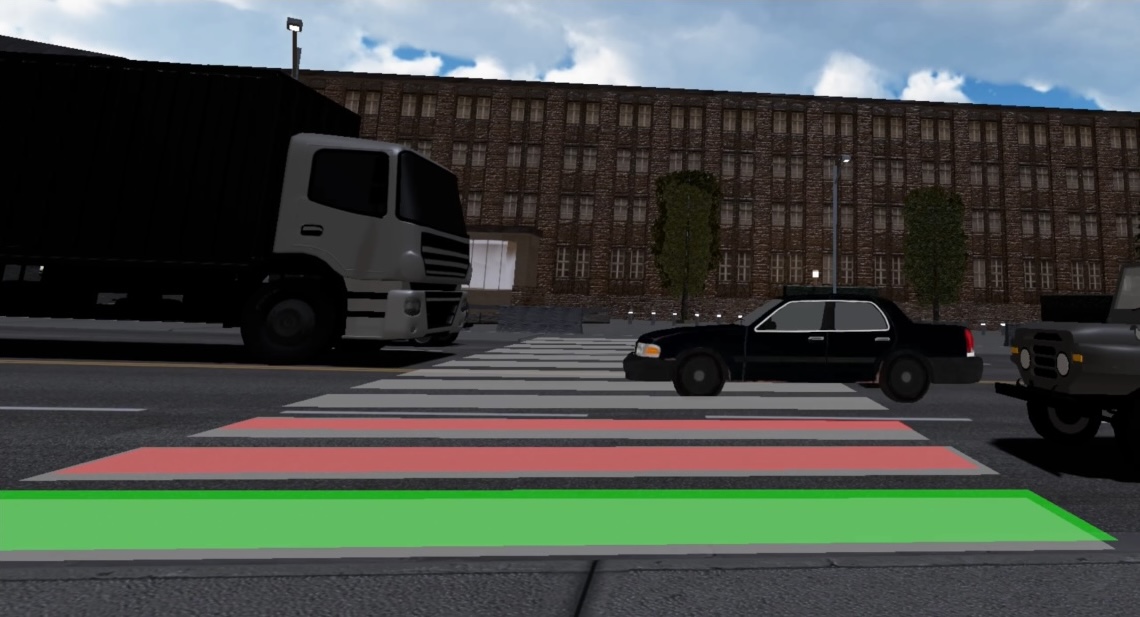}
        \caption{\centering  eHMI proposed in this study\\ convey ego's intent and surrounding risk}
        \label{fig:proposed_ehmi}
    \end{subfigure}
    \vspace{0.6mm}
    \caption{Overview of the VR user study. Left: Baseline without eHMI. Middle: Standard ego's intent-only eHMI signals the pickup's yielding intent, which potentially causes pedestrians to overlook surrounding hazards. Right: Proposed Attention-Guiding eHMI (AGeHMI). This interface not only signals the pickup's yielding intent (green bar closest to the pedestrian) but also employs directional projections to visualize surrounding risks (two red bars further away), effectively guiding pedestrian attention toward hazards.}

    \label{fig:design}
\end{figure*}
\vspace{-2mm}

With the rapid development of autonomous driving technology, vehicles are evolving from manual driving modes to highly or fully automated forms, bringing higher driving efficiency and potential safety improvements to the transportation system \cite{FAGNANT2015167}. However, the proliferation of autonomous vehicles also introduces new interaction challenges: non-verbal communication methods commonly seen in traditional driving contexts, such as eye contact and gestures, disappear as drivers are removed from the control loop, making it difficult for vulnerable road users (VRUs) like pedestrians and cyclists to discern the vehicle's intentions and attention \cite{8667866}. 

Accordingly, the studies of external Human-Machine Interface (eHMI) have emerged. Typically installed on the exterior of the vehicle, eHMI aims to convey the vehicle's status and driving intentions to nearby pedestrians or other traffic participants, thus enhancing pedestrians' trust in autonomous vehicles and reducing the likelihood of traffic conflicts and accidents \cite{eHMI_survey,eHMI_review}.
Visual eHMI is currently the most widely researched category, encompassing textual or symbolic displays \cite{eHMI_display,EISELE20221}, LED light bands \cite{Nianzhao_CHI24,Nianzhao_ACCESS25}, road projections \cite{10071038,autoUI19,autoUI19}, and anthropomorphic designs incorporating gestures \cite{eHMI_arm}, movable parts like "eyes" \cite{eHMI_eye}, or facial expressions \cite{deClercq2019}.
Auditory eHMIs have also been investigated to compensate for visual limitations, utilizing external alert tones \cite{10.1007/978-3-030-78645-8_27} or voice announcements \cite{eHMI_sound} to signal vehicle movements.
Beyond single-channel approaches, multimodal strategies combining visual cues with auditory or tactile feedback have been proposed to accommodate diverse pedestrian needs and environmental conditions \cite{s22124537,s21165274, eHMI_CHI25}. 

However, as highlighted in multiple studies, existing eHMI designs often induce unintended behavioral deficits in pedestrians \cite{deClercq2019,s22093339}. 
Specifically, two critical issues arise. First, eHMI-equipped vehicles tend to capture excessive visual attention, causing pedestrians to reduce their scanning of other environmental hazards (often referred to as “tunnel vision”). Second, although most eHMIs are designed to communicate only the ego-vehicle's intent, pedestrians frequently misinterpret this as a safety clearance. They may erroneously assume the AV is monitoring the surroundings on their behalf, leading to over-reliance and causing them to cross without independently verifying traffic conditions.

To address these challenges, our work makes three key contributions. First, we propose the Attention-Guiding eHMI (AGeHMI), a new design concept of eHMI. Unlike existing designs that solely convey the ego-vehicle's status, AGeHMI aims to actively guide VRUs' attention toward potential environmental hazards. 
Second, we instantiate this concept through a projection-based visualization design that intuitively maps environmental hazards directly onto the road surface. By utilizing directional cues and risk-based color coding, this design enables pedestrians to rapidly perceive surrounding traffic threats. 
Finally, we validate our approach through a user study (N = 20) in Virtual Reality (VR), providing empirical evidence of its safety and usability benefits. The results demonstrate that AGeHMI significantly influences participants' visual attention distribution and reduces collision risks with surrounding vehicles, while simultaneously improving confidence and minimizing cognitive workload.

\section{Designing Projection-based Attention-Guiding eHMI}
\label{sec: proposal}

Following Fig.\ref{fig:design_scenario}, this section introduces AGeHMI and its projection-based instantiation. We adopt a projection-based approach because it is the most practical and widely implemented eHMI modality in real-world deployments \cite{audi_a6_e_tron,gongjiaotisheng,author2025title}.

\vspace{-2mm}
\begin{figure*}[h!]
    \centering
    % Right Figure
        \centering
        \includegraphics[width=0.5\linewidth]{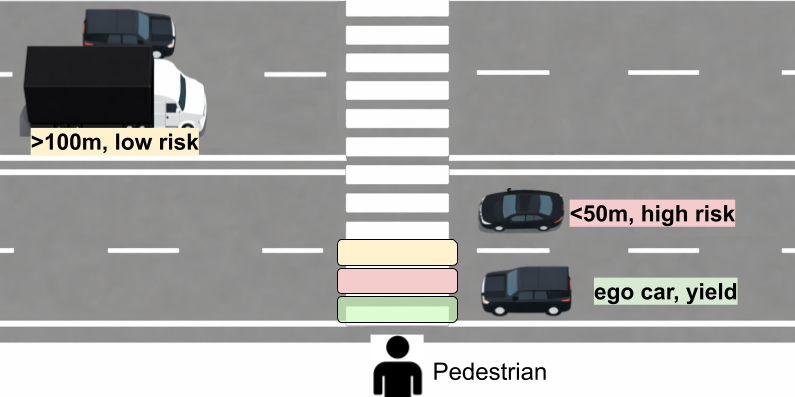}
        \caption{\centering Projection-based AGeHMI, illustrating the depth-based spatial mapping of hazards and the risk-based color coding.}
        \label{fig:design_scenario}
    % \vspace{-2mm}
\end{figure*}
\vspace{-2mm}

The AGeHMI design involves two key dimensions of information encoding: spatial mapping and risk visualization.
The first dimension is the spatial mapping of hazard sources. The primary goal is to ensure immediate geometric intuitiveness by mapping the physical layout of the road lanes directly onto the projection interface. We define the projection area as the reference frame:
(i) The bars projected closest to the pedestrian correspond to the closest lane (the ego-vehicle's lane);
(ii) The bars projected further away correspond to the spatially more distant lanes (e.g., opposing traffic lanes or adjacent same-direction lanes).
This depth-based mapping allows pedestrians to instinctively link a specific projected cue to the corresponding physical location in the real world without needing to mentally translate the information.

The second dimension is the visualization of risk levels through color coding. While the position of the projection indicates the lane, the color indicates the urgency of the threat within that lane. We employ a progressive color scheme:
(i) green indicates that it is safe to proceed or that the area poses low risk;
(ii) yellow indicates that moderate caution is required;
(iii) red denotes a high-risk zone requiring immediate attention.
From the pedestrian’s perspective, this design is immediately intuitive: the projected colors act as direct attentional cues, with each color corresponding to a different level of risk.
From a system implementation perspective, the underlying risk assessment logic differs between the ego vehicle and the surrounding traffic. For the ego vehicle, the assessment is deterministic, as its own motion planning and intent are explicitly known. Consequently, the color coding directly reflects its intended behavior: green signifies a yielding intention (implying a safe state for pedestrians), while red indicates a non-yielding state (implying high risk).
For surrounding vehicles, the system relies on external perception sources, such as on-board sensor measurements or Vehicle-to-Vehicle (V2V) communication, to estimate threat levels. In our study, we employed a distance-based heuristic: a lane was classified as low risk if no vehicle was present, moderate risk if a vehicle was detected within 100 meters of the crosswalk, and high risk if a vehicle was within 50 meters.

Additionally, there are two points to note:
First, the spatial mapping of hazard sources is limited to depth only, in order to minimize cognitive load and visual clutter; more detailed lateral directionality within a single lane is not considered. For example, even if a vehicle is approaching from the left, the system does not split the corresponding bar into different colors (e.g., red on the left, yellow on the right).
Second, the encoding of risk level should follow a "conservative safety" strategy. In situations of uncertainty where the autonomous vehicle cannot definitively confirm that a surrounding area is safe, the system defaults to visualizing it as a potential risk to avoid providing false reassurance.

\section{VR User Study}
\label{sec: method}
\subsection{Study Setup}

\textbf{Task.}
Referring to Fig. \ref{fig:design}, participants wearing the Meta Quest 3 head-mounted display will be positioned on a virtual sidewalk at a pedestrian crossing (developed using Unity), under three different eHMI conditions (No eHMI, Ego-only eHMI, AGeHMI). Their primary task was to use a `feel-safe'' button to indicate, in real time, whether they felt it was safe to cross \cite{10.1145/3750069.3750110, deClercq2019}. Specifically, participants were instructed to: (i) press and hold the button of the Quest3 standard controller whenever they felt safe to cross; (ii) keep the button pressed as long as they continued to feel safe; (iii) release the button the moment they no longer felt safe. 

\textbf{Participants.}
We recruited 20 participants (8 female) from a local university, aged between 19 and 41 years ($M=27.10$, $SD=7.25$), all with normal or corrected-to-normal vision. 14 participants held a valid driver’s license, and reported a mean driving experience of 5.25 years ($SD=6.97$). Regarding technology familiarity, 16 participants had prior VR experience, and self-rated knowledge of autonomous vehicles averaged 3.35 out of 5 ($SD=0.99$).

\textbf{Vehicle configurations and behaviors.}
As shown in Fig. \ref{fig:design_scenario}, we consider four vehicle types: (i) Ego vehicle: the autonomous vehicle; (ii) Small vehicle: partially obscured by the ego vehicle; (iii) Opposing large truck: traveling in the lane facing the participant; (iv) Small vehicle: behind the large truck. 
We note that it is a unique aspect of our study to incorporate such a realistic and complex multi-vehicle scenario, including obscured vehicles with potential occlusion.
All vehicles followed standardized behaviors \cite{Deligianni2017AnalyzingAM,Nianzhao_CHI24}: yielding vehicles decelerated at $-2.4\,\text{m/s}^2$ from 45\,m, stopped at 5\,m before the crosswalk, paused for 5\,s, then accelerated away. Non-yielding vehicles maintained a constant speed of 50\,km/h from 100\,m, passing through the crosswalk without stopping.

\subsection{Measurements and Data Processing}
For the objective metrics, we measured two outcomes:
(i) \textit{potential collision rate (PCR)}, defined as the proportion of trials in which a participant’s “feel-safe” button press temporally overlapped with any vehicle passing through the pedestrian crosswalk. This metric aims to quantify potential safety failures, such as instances where participants mistakenly identified a non-yielding vehicle as yielding, or failed to notice an approaching non-yielding vehicle entirely, leading to risky crossing behavior; and (ii) \textit{visual attention distribution}, i.e.,  the respective time participants spent visually focusing on (or directing their gaze toward) each of the four vehicles, to quantify the effectiveness of the eHMI in guiding visual attention. Given that the Quest 3 does not support native eye-tracking, we relied on head tracking instead. Furthermore, considering that the field of view (FOV) in Quest 3 is narrower than that of natural human vision, we defined a vehicle as "observed" whenever it was rendered within the active VR display. Furthermore, to ensure data relevance and minimize behavioral noise (e.g., scanning the stopping vehicle without purpose), we restricted our analysis to the first 10 seconds of each trial, which is the critical phase for situational awareness and crossing decision-making.

For subjective metrics, we collected subjective feedback using a 7-point Likert scale focusing on two primary dimensions: (i) \textit{cognitive load}, assessed via the standard NASA-TLX sub-scales (mental, physical, and temporal demand; performance; effort; and frustration) \cite{hart1988nasatlx}, and (ii) \textit{user experience and acceptance}, which covered perceived usefulness, perceived ease of use, trust, social acceptance, and overall evaluation. Given the ordinal Likert data and our sample size ($N=20$), we used non-parametric statistics: Friedman tests for main effects, followed by Wilcoxon signed-rank tests with Bonferroni correction ($\alpha \approx 0.017$) for post-hoc comparisons.

\subsection{Procedure}
Before the experiment began, participants signed a written informed consent form and received an overview of the study’s purpose and procedures.\footnote{The study was granted an exemption by the Institutional Review Board (IRB) of the affiliated institution. IRB details are omitted for double-blind review.} They were then guided to complete a brief training session to familiarize themselves with the VR environment and the ``feel-safe'' button task, which included 10 randomized practice trials.
After training, participants proceeded to the main experiment. The study followed a within-subjects design with three blocks, corresponding to the three eHMI conditions. The order of these blocks was counterbalanced across participants (using a Latin Square design) to minimize learning and order effects. 
Within each block, participants experienced 22 randomized vehicle configurations, varying both the presence of each vehicle and whether it would yield or not. Upon completing the 22 scenarios for a specific eHMI condition, participants completed the subjective questionnaires for that specific eHMI condition and took a short break to minimize VR fatigue.
After completing all three blocks, a semi-structured interview was conducted to gather qualitative feedback, focusing on overall impressions, potential misunderstandings of the projection signals, and specific eHMI condition features that influenced their perceived safety and decision-making. The entire user study lasted approximately 40 minutes.

\section{Results and Discussion}

\subsection{Results}
% \vspace{-1mm}
\begin{figure*}[th!]
    \centering
    % --- 第 1 个图 ---
    \begin{minipage}[t]{0.32\textwidth}
        \centering
        \includegraphics[width=\linewidth]{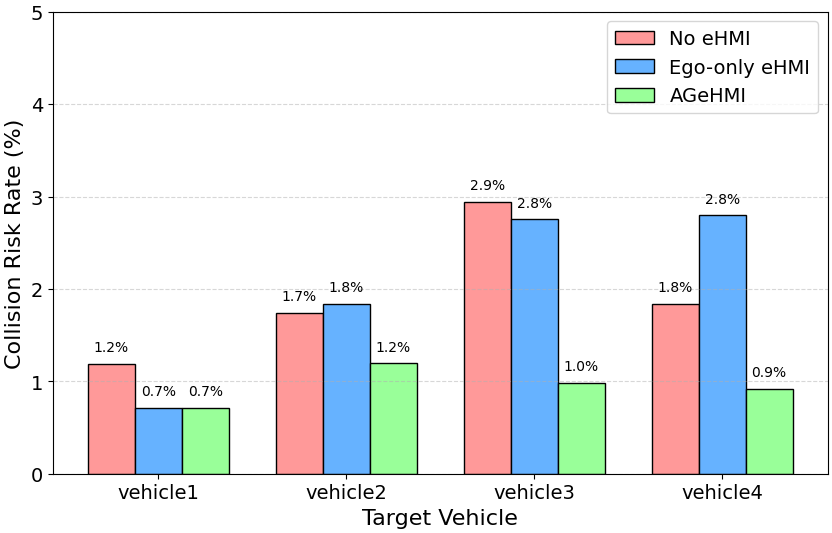} 
        \caption{\centering Comparison of potential collision rates for each target vehicle}
        \label{fig:collision_rate}
    \end{minipage}
    \hfill % 这是一个弹性空格，用于撑开间距
    \begin{minipage}[t]{0.32\textwidth}
        \centering
        % 图片宽度设置为 \linewidth 以适应 minipage 的大小
        \includegraphics[width=\linewidth]{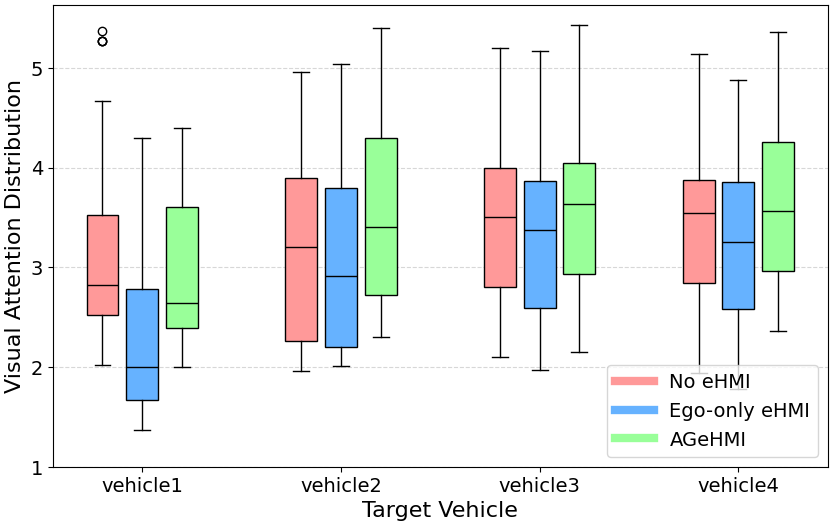}
        \caption{\centering Distribution of visual attention for each target vehicle}
        \label{fig:vehicle_observation}
    \end{minipage}
    \hfill % 这是一个弹性空格
    \begin{minipage}[t]{0.32\textwidth}
        \centering
        \includegraphics[width=\linewidth]{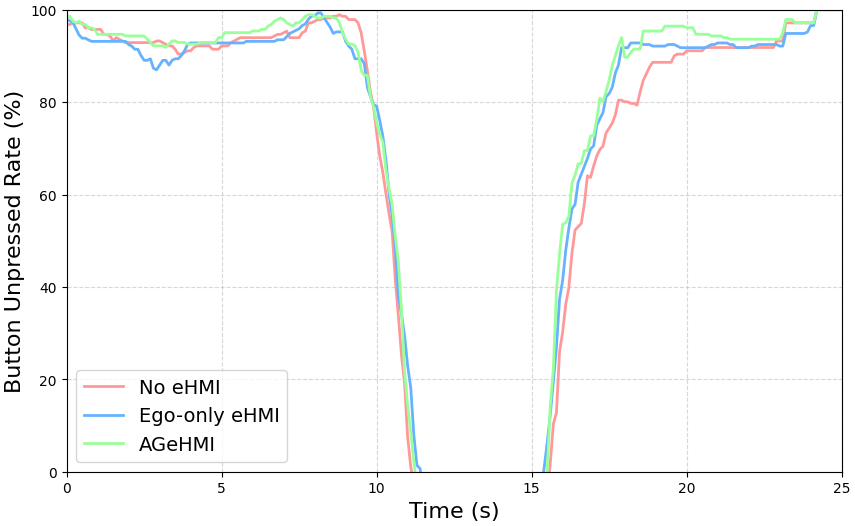}
        \caption{\centering Temporal percentage of button unpresses over time}
        \label{fig:button_probability}
    \end{minipage}
    \label{fig:collision_risk}
    % \vspace{-2mm}
\end{figure*}
\vspace{-2mm}

Figures \ref{fig:collision_rate}-\ref{fig:button_probability} present the collision risk rates for various vehicles, suggesting the potential safety advantages of AGeHMI. For the ego vehicle (Vehicle 1), both \textit{AGeHMI} and \textit{Ego-only} achieved similarly low risk levels (0.7\%), indicating comparable effectiveness in intent communication. However, for the surrounding vehicles (2–4), \textit{AGeHMI} kept risks below 1.2\%, while other methods reached up to 2.9\%. Notably, for the potentially occluded Vehicle 4, the collision rate with \textit{Ego-only} (2.8\%) was actually higher than with \textit{No eHMI} (1.8\%), suggesting that the \textit{Ego-only} condition may in fact increase risk in complex scenarios by fostering a false sense of security regarding hidden hazards.

To clarify the behavioral factors underlying these collision risks, Figure \ref{fig:vehicle_observation} examines visual attention distribution. Under the \textit{Ego-only} condition, participants paid less attention to surrounding vehicles, echoing prior findings of “tunnel vision.” In contrast, \textit{AGeHMI} resulted in longer gaze times for all surrounding vehicles, confirming that its cues encouraged broader hazard checking. Surprisingly, although \textit{AGeHMI} clearly conveyed the ego vehicle’s intention, the visual attention duration for the ego vehicle itself was significantly higher than in the \textit{Ego-only} condition. User interviews clarified this behavior: participants reported that they perceived the projections as a unified display, “Although the color semantics differ for the ego vehicle and surrounding vehicles, I intuitively process them in the same way.”

Figure \ref{fig:button_probability} further investigates decision timing. Across all conditions, the sense of safety peaked (i.e., the proportion of button presses reached its maximum) as the yielding vehicle came to a stop (11–15 s). However, \textit{AGeHMI} kept participants more cautious during the initial phase (0–8 s), with a higher rate of unpressed buttons. \textit{Ego-only} exhibited a more pronounced early drop and subsequent rebound, suggesting premature and wrong decisions that were later corrected. After the yielding vehicles resumed movement (>15 s), participants in the \textit{AGeHMI} condition released the button more quickly, potentially indicating that \textit{AGeHMI} enhanced situational awareness, enabling participants to more rapidly recognize renewed risks and respond accordingly.

\vspace{-1mm}
\begin{figure*}[th!]
    \centering
    % 左图 (图4) - 按照像素比例分配约为 53% 的宽度
    \begin{minipage}[t]{0.53\textwidth}
        \centering
        \includegraphics[width=\linewidth]{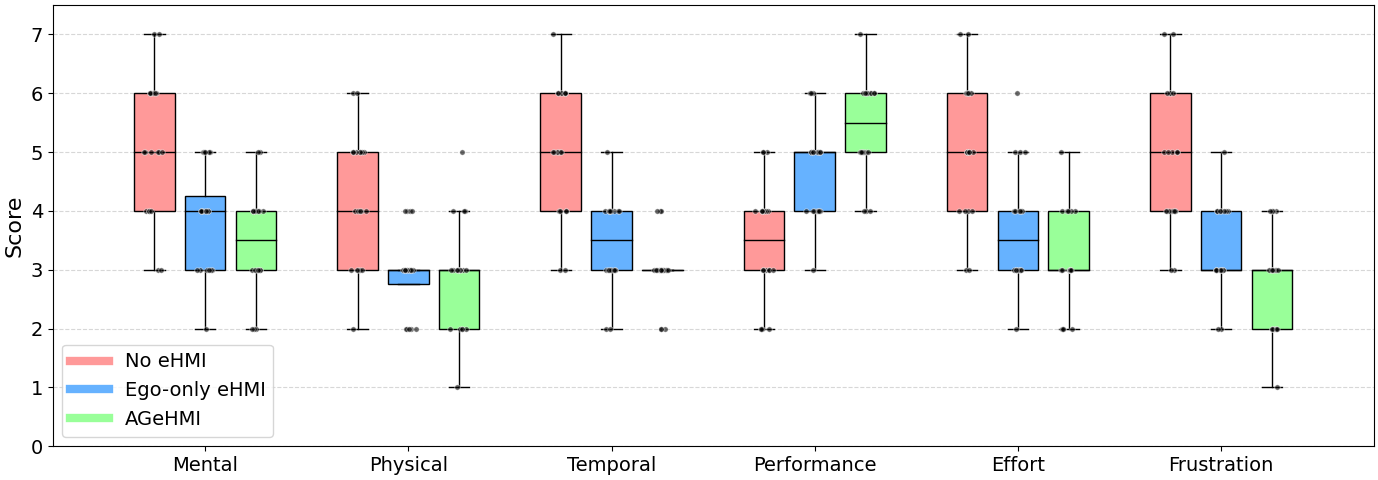}
        \vspace{-1mm}
        \caption{\centering Boxplots of NASA-TLX scores, with dots representing individual data points. Lower scores indicate better, except for Performance.}
        \label{fig:nasa_tlx}
    \end{minipage}%
    \hfill % 在两图之间填充空白，使其撑满行宽
    % 右图 (图5) - 按照像素比例分配约为 45% 的宽度
    \begin{minipage}[t]{0.45\textwidth}
        \centering
        \includegraphics[width=\linewidth]{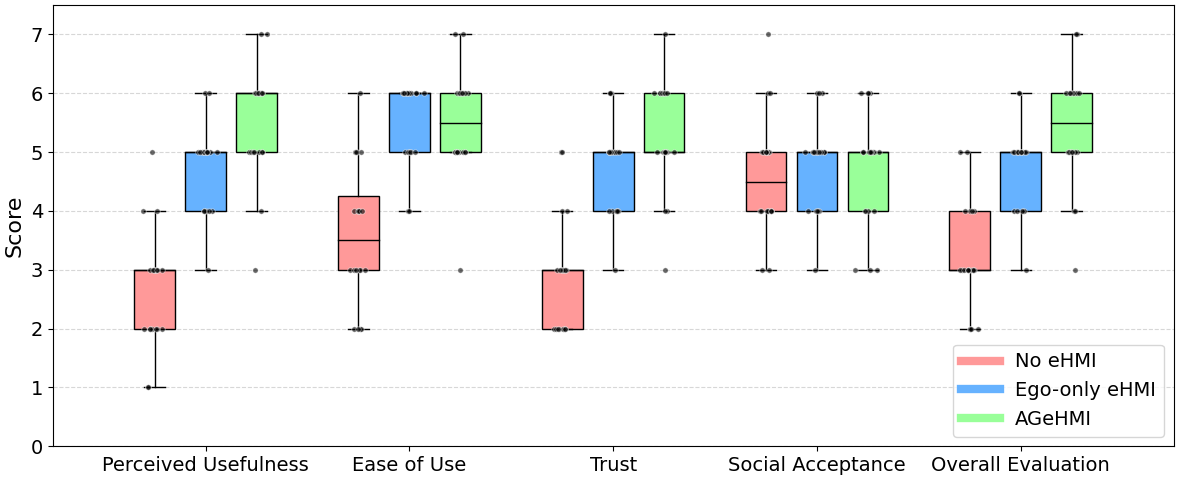} 
        \vspace{-1mm}
        \caption{\centering Boxplots of User Experience Metrics. Higher scores indicate better outcomes.}
        \label{fig:ux_ratings}
    \end{minipage}
    \vspace{-1mm}
\end{figure*}
\vspace{-1mm}

Figure~\ref{fig:nasa_tlx} summarizes cognitive efficiency and workload results. Participants rated their performance higher with \textit{AGeHMI} ($M=5.40$) than with \textit{Ego-only} ($M=4.70$, $p < .001$), and reported lower frustration ($M=2.75$ vs. $M=3.35$, $p < .001$). AGeHMI also lowered temporal demand ($M=3.00$ vs. $M=3.45$, $p = .003$), enabling calmer, less pressured decisions. These benefits came without added cognitive load: mental demand for AGeHMI ($M=3.45$) was similar to Ego-only ($M=3.85$, $p = .130$), and physical demand  (e.g., turning the body or head to check for vehicles) remained low ($p = .78$). Although effort ratings favored AGeHMI ($M=3.15$ vs. $M=3.70$), this was not statistically significant ($p = .039$). User feedback indicated that ``while AGeHMI causes an additional one-time demand in understanding the cues, it also reduced the effort required for subsequent situational awareness and enhanced confidence.''

For user experience and acceptance (Figure~\ref{fig:ux_ratings}), AGeHMI led in perceived usefulness ($M=5.50$) and trust ($M=5.30$), both significantly higher than Ego-only ($p < .005$). Participants described this trust as informed rather than blind reliance, leading to a significantly higher overall evaluation preference ($M=5.40$ vs. $M=4.75, p < .001$).
Perceived ease of use was equal for both eHMIs ($M=5.50$, $p = .86$), though participant opinions varied: participants who favored the \textit{Ego-only} emphasized its minimalist nature, noting that ``it just tells me one thing, which is simple to process.'' Conversely, proponents of AGeHMI argued that the rich visualization actually made it easier to use by externalizing the risk assessment.
For social norms, AGeHMI’s rating ($M=4.70$) was similar to Ego-only and No eHMI ($p = .79$). Supporters of \textit{No eHMI} worried about “semantic interference” with existing traffic signals and light pollution. Supporters of Ego-only saw it as a natural upgrade to current signals (e.g., advanced brake lights), while some favored AGeHMI as dynamic infrastructure for future, unsignalized environments (e.g., eHMI as a traffic signal in unsignalized areas).

\subsection{Discussion and Limitations}

First, regarding the experimental design, the use of VR limits ecological validity, as it lacks real-world physical risks and environmental variables such as harsh sunlight that could affect projection visibility. Future studies should validate these findings in real-world test tracks or by using AR under varying lighting conditions to assess physical feasibility. Additionally, our sample consisted primarily of young students and middle-aged faculty, which limits the generalizability of our findings across different age groups and cultural backgrounds. Expanding the participant pool to include children, the elderly, and individuals from diverse cultural backgrounds is necessary to evaluate the universal comprehensibility.

Second, with respect to the specific projection-based eHMI design, scalability remains a challenge where multiple projecting vehicles could cause visual clutter and confusion. Therefore, implementing projection coordination via V2V communication would be a future direction. Furthermore, reliance on color coding may present accessibility barriers for color-blind users and could potentially conflict with existing traffic infrastructure (e.g., a green projection appearing alongside a red traffic light). To address this, future iterations should consider more diverse and inclusive designs, as well as more context-aware solutions (e.g., complementary to traffic lights) for eHMI.

Third, there are fundamental challenges inherent to the AGeHMI concept itself. One issue is sensor reliability; false negatives in detecting occluded vehicles could result in incorrect or incomplete cues being conveyed. Integrating V2X technology is a potential direction to enhance perception capabilities beyond on-vehicle sensors and ensure data accuracy. Similarly, users may misinterpret attention guidance as a guarantee of safety, leading to legal and ethical concerns. Addressing this requires long-term public education to cultivate appropriate mental models, as well as interface designs that clearly distinguish between ``suggestions'' and authoritative ``instructions'' to prevent over-reliance.

\section{Conclusion and Future Work}
\label{sec: conclusion}
In this paper, we addressed a critical paradox: while traditional eHMIs effectively communicate the yielding intention of a single autonomous vehicle, they often cause pedestrians to neglect surrounding environmental hazards. To address this, we proposed the concept of AGeHMI and instantiated the concept with a projection-based design by visualizing the risks of surrounding traffic.
Our results demonstrate that, from an objective perspective, AGeHMI significantly reduced collision risks by effectively guiding visual attention toward hazards. Subjectively, participants rated the design as highly trustworthy and useful, reported enhanced confidence, and experienced reduced cognitive workload.

In future work, we plan to further refine the AGeHMI concept. 
First, we will assess the impact of visual conspicuity factors, such as vehicle body color (e.g., white vs. dark vehicles) and road texture, to optimize the contrast and legibility of the projections across diverse real-world environments. 
Second, we intend to expand the scope of our user studies to include a larger and more culturally diverse participant pool, thereby validating the generalizability of our findings and exploring how varying traffic norms influence the interpretation of risk-based projections.

\bibliographystyle{ACM-Reference-Format}
% \balance
\bibliography{bib}
\end{document}